%% file: main.tex
\documentclass[sigconf,nonacm]{acmart}

\usepackage{subfigure}
\usepackage{siunitx}
\usepackage{enumitem}
\usepackage{xurl}

\usepackage{algorithm}
\usepackage[noend]{algpseudocode}
\let\ReturnInline\Return
\renewcommand{\Return}{\State\ReturnInline}
\algrenewcommand\algorithmicrequire{$\rhd$}
\algrenewcommand\algorithmicensure{$\square$}

\AtBeginDocument{%
  \providecommand\BibTeX{{%
    \normalfont B\kern-0.5em{\scshape i\kern-0.25em b}\kern-0.8em\TeX}}}

\setcopyright{acmcopyright}
\copyrightyear{2018}
\acmYear{2018}
\acmDOI{XXXXXXX.XXXXXXX}

\acmConference[Conference acronym 'XX]{Make sure to enter the correct
  conference title from your rights confirmation emai}{June 03--05,
  2018}{Woodstock, NY}

\newcommand{\ignore}[1]{}

\begin{document}

\title{Heuristics for Inequality minimization in PageRank values}


\author{Subhajit Sahu}
\email{subhajit.sahu@research.iiit.ac.in}
\affiliation{%
  \institution{IIIT Hyderabad}
  \streetaddress{Professor CR Rao Rd, Gachibowli}
  \city{Hyderabad}
  \state{Telangana}
  \country{India}
  \postcode{500032}
}


\settopmatter{printfolios=true}

\begin{abstract}
PageRank is a widely used algorithm for ranking webpages and plays a significant role in determining web traffic. This study employs the Gini coefficient, a measure of income/wealth inequality, to assess the inequality in PageRank distributions and explores six deterministic methods for reducing inequality. Our findings indicate that a combination of two distinct heuristics may present an effective strategy for minimizing inequality.
\end{abstract}




\maketitle

\section{Introduction}
\label{sec:introduction}
\input{01-introduction}

\section{Related work}
\label{sec:related}
\input{02-related-work}

\section{Preliminaries}
\label{sec:preliminaries}
\input{03-preliminaries}

\section{Approach}
\label{sec:approach}
\input{04-approach}

\section{Evaluation}
\label{sec:evaluation}
\input{05-evaluation}

\section{Conclusion}
\label{sec:conclusion}
\input{06-conclusion}

\begin{acks}
I would like to thank Prof. Kishore Kothapalli, Prof. Hemalatha Eedi, and Prof. Sathya Peri for their support.
\end{acks}

\bibliographystyle{ACM-Reference-Format}
\bibliography{main}

\end{document}

%% file: 01-introduction.tex
The PageRank algorithm is a critical tool in web search and ranking, determining the order of search results on popular search engines. This algorithm evaluates webpages' popularity based on the idea that pages linked by other popular pages are themselves considered popular. However, this can result in a positive feedback loop, where already well-linked/popular pages receive even more traffic, further marginalizing less-connected nodes and exacerbating inequality.

Addressing inequality is essential for improving the outcomes of PageRank-based analyses. Extreme imbalances can worsen disparities, leading to negative consequences in various fields. In social networks, reducing inequality in PageRank values helps prevent the underrepresentation of certain groups. In information retrieval, minimizing ranking inequality prevents dominance by a few sources. In supply chains, PageRank-like algorithms identify critical suppliers. Excessive favoritism towards a few can create unhealthy dependencies, risking disruptions from events like political instability or natural disasters. In financial systems, addressing inequality in PageRank rankings allows smaller institutions to access capital, promoting innovation and competition. Less concentration in rankings reduces vulnerability to cascading failures, as shocks are less likely to spread uncontrollably. In military networks, minimizing PageRank inequality reduces over-reliance on a few central nodes, enhancing resilience to targeted attacks and ensuring decentralized communication and decision-making. In communication and data networks, less centralization mitigates risks from cyberattacks or natural disasters by preventing reliance on a few data centers.

This short paper explores various heuristics aimed at minimizing inequality in PageRank values.

%% file: 02-related-work.tex
Algorithmic fairness has become a significant area of focus in recent research \cite{tsioutsiouliklis2021fairness}. A network is deemed fair when nodes from two distinct groups occupy equally central positions \cite{pitoura2023pagerank}. Structural biases present in social networks can influence the fairness of algorithms used for their analysis \cite{saxena2022fairsna}. Xie et al. \cite{xie2021fairrankvis} developed a framework to visualize multi-class bias in graph algorithms. Karimi et al. \cite{karimi2018homophily} demonstrated that minority nodes in homophilic networks face challenges in achieving higher degree ranks due to disparities in community sizes. Tsioutsiouliklis et al. \cite{tsioutsiouliklis2021fairness} introduced fairness-sensitive and locally fair PageRank algorithms and proposed the concept of universal personalized fairness. Krasanakis et al. \cite{krasanakis2021applying} proposed a fair ranking algorithm with personalization that maintains rank quality despite biased inputs. Link recommendation methods optimized for fairness were suggested by Tsioutsiouliklis et al. \cite{tsioutsiouliklis2022link}. However, our work diverges slightly from these studies, as it focuses on minimizing absolute inequality in PageRank scores, with inequality measured using the Gini coefficient \cite{gini1912italian}.

Addressing inequality is crucial for ensuring that solutions are acceptable and effective across various domains \cite{fox1966discrete, dyer1977note, hung2006allocation}. From public services to disaster relief, reducing inequality can mitigate dissatisfaction and improve societal outcomes \cite{yang2013call, gutjahr2018equity}. However, the complexity of inequality modeling lies in the context-specific nature of its definitions, requiring tailored approaches to address the unique challenges of each application \cite{shehadeh2023equity}.


%% file: 03-preliminaries.tex
\subsection{PageRank}
\label{sec:PageRank}

Consider a directed graph $G(V, E, w)$, with $V$ ($n = |V|$) as the set of vertices and $E$ ($m = |E|$) as the set of edges. The PageRank $R[v]$ of a vertex $v \in V$ in this graph measures its importance based on incoming links and their significance \cite{rank-page99}. Equation \ref{eq:pr} defines the PageRank calculation for a vertex $v$ in $G$. $G.in(v)$ and $G.out(v)$ represent incoming and outgoing neighbors of $v$, and $\alpha$ is the damping factor (usually $0.85$). Initially, each vertex has a PageRank of $1/n$, and the \textit{power-iteration} method updates these values iteratively until they converge within a specified tolerance $\tau$, indicating that convergence has been achieved.

\begin{equation}
\label{eq:pr}
    R[v] = \alpha \times \sum_{u \in G.in(v)} \frac{R[u]}{|G.out(u)|} + \frac{1 - \alpha}{n}
\end{equation}

\subsection{Gini coefficient}

Gini coefficient $G$ \cite{gini1912italian} is a value which represents income/wealth inequality within a nation or group. It ranges from $0$ to $1$, with $0$ representing total equality and $1$ representing total inequality. It is calculated from the Lorenz curve, which plots cumulative income/wealth against cumulative number of households/people. It is calculated using Equation \ref{eq:gini}, where $A$ is the area between the line of perfect equality and the Lorenz curve, and $B$ is the total area under the line of perfect equality.

\begin{equation}
\label{eq:gini}
  G = \frac{A}{A+B}
\end{equation}

%% file: 04-approach.tex
We now study minimization of Gini coefficient of PageRank values on a number of graphs, using six different deterministic heuristics for adding edges to the graph. First, the PageRank of each vertex is computed in the original graph, and the original Gini coefficient is obtained. A heuristic is then run to obtain the most suitable edge to be added. After this edge is added, the same heuristic is run again. For each heuristic $1000$ edges are added. We plot the variation of Gini coefficient with each added edge for each heuristic.

Our first heuristic, \textbf{Cxrx}, adds an edge between the highest contributing vertex to the lowest rank vertex. The idea behind this heuristic is to provide the highest possible increase in rank to the lowest rank vertex. We obtained the highest contributing vertex by finding the vertex with highest $R/(d+1)$ value.

The second heuristic, which we refer to as \textbf{CxSx}, is based on the idea of providing the highest possible increase in rank to a vertex which directly or indirectly links to many other vertices (so that it increases the rank of a large number of other vertices as well). This is achieved by adding an edge from the highest contributing vertex to the vertex with highest reverse PageRank. Here, the reverse PageRank of a vertex is obtained by reversing (transposing) the graph, and calculating the PageRanks.

The third heuristic called \textbf{CxSr} is an extension of \textbf{CxSx}, and it prioritizes increasing the rank of vertices which link (directly or indirectly) to a large number of vertices having a low PageRank score. This is done by calculating a modified reverse PageRank, that prioritizes contribution from vertices with low forward PageRank. Here, the reverse rank of each vertex is calculated as $r_u = \alpha (1 - R_u) r_v / d_v + (1-\alpha)/N$, where $r_u$ is the reverse rank of a given vertex and $R_u$ is its forward rank (precomputed), $r_v$ is the reverse rank of a target vertex and $d_v$ is its in-degree, $\alpha$ is the damping factor, and $N$ is the number of vertices in the graph.

The remaining three heuristics \textbf{CRrx}, \textbf{CRSx}, and \textbf{CRSr} are a variation of the three heuristics mentioned above where the source vertex is chosen such that it minimizes the rank of the highest ranked vertex. That is, we choose the source vertex with highest contribution to the highest rank vertex. The idea is to reduce rank of high-ranked vertices and increase the rank of low-ranked vertices at the same time, thus reducing inequality.

Algorithm \ref{alg:heuristic} presents the pseudocode for the six proposed heuristics. It accepts as input a graph $G$, a heuristic type $\eta$, and a number $\Delta$ that indicates how many edges to add. For each edge addition (up to $\Delta$ times), the algorithm selects a source vertex $u$ and a target vertex $v$ based on the chosen heuristic. This heuristic may prioritize nodes with high graph contributions or those connected to higher-ranked nodes. The output is a modified graph $G$ with a more balanced PageRank distribution.

\input{src/alg-heuristic}

%% file: src/alg-heuristic.tex
\begin{algorithm}[hbtp]
\caption{Minimizing inequality in PageRank using Heuristics.}
\label{alg:heuristic}
\begin{algorithmic}[1]
\Require{$G(V, E)$: Input graph}
\Require{$\eta$: Heuristic type (such as 'Cxrx', 'CxSx')}
\Require{$\Delta$: Number of edges to add}
\Ensure{$G'$: Transpose of input graph}
\Ensure{$F$: Rank scaling factor}
\Ensure{$\alpha$: Damping factor}

\Statex

\Function{minimizePagerankInequality}{$G, \eta, \Delta$}
  \ForAll{$l_i \in [0\ \dots\ \Delta)$}
    \State $(u, v) \gets selectEdge(G, \eta)$
    \State $G.addEdge(u, v)$
  \EndFor
  \Return{G}
\EndFunction

\Statex

\State $\rhd$ Select an edge to add to the graph,
\State $\rhd$ based on the chosen heuristic
\Function{selectEdge}{$G, \eta$}
  \State $R \gets ranks(G)$
  \State $R' \gets ranks(G')$
  \State $R'' \gets ranks(G', 1 - R)$
  \If{$\eta = Cx**$}
    \State $u \gets maxContributing(G, R)$
  \Else
    \State $u \gets maxContributingToHighRank(G, R)$
  \EndIf
  \If{$\eta = **rx$}
    \State $v \gets \arg \min\ \{R[x] : x \in V\}$
  \ElsIf{$\eta = **Sx$}
    \State $v \gets \arg \max\ \{R'[x] : x \in V\}$
  \Else
    \State $v \gets \arg \max\ \{R''[x] : x \in V\}$
  \EndIf
  \Return{$(u, v)$}
\EndFunction

\Statex

\State $\rhd$ Choose source vertex with highest contribution
\Function{maxContributing}{$G, R$}
  \State $v \gets \arg \max\ \{R[x] / |G.out(x)| : x \in V\}$
  \Return{$v$}
\EndFunction

\Statex

\State $\rhd$ Choose source vertex with highest
\State $\rhd$ contribution to the highest rank vertex
\Function{maxContributingToHighRank}{$G, R$}
  \State $v \gets \arg \max\ \{R[x] : x \in V\}$
  \State $u \gets \arg \max\ \{R[x] / |G.out(x)| : x \in G.neighbors(v)\}$
  \Return{$u$}
\EndFunction

\Statex

\State $\rhd$ Compute PageRank scores with scaling factor $F$
\Function{ranks}{$G, F = 1$}
  \State $R \gets \{1/|V|\ : x \in V\}$
  \While{\textbf{not converged}}
    \ForAll{$u \in V$}
      \State $R[u] \gets \alpha F[u] R[v] / |G.out(v)| + (1-\alpha)/|V|$
    \EndFor
  \EndWhile
\EndFunction
\end{algorithmic}
\end{algorithm}

%% file: 05-evaluation.tex
\subsection{Experimental Setup}
\label{sec:setup}

\subsubsection{System used}

We employ a server consisting of two Intel Xeon Gold 6226R processors, each featuring $16$ cores running at $2.90$ GHz. Each core includes a $1$ MB L1 cache, a $16$ MB L2 cache, and shares a $22$ MB L3 cache. The system is equipped with $376$ GB of RAM and runs CentOS Stream 8.

\subsubsection{Configuration}

We utilize 32-bit integers for representing vertex IDs and 32-bit floats for edge weights, while computations and hashtable values are based on 64-bit floats. We use $64$ threads by default, which to correspond to the available cores on the system. Compilation is conducted using GCC 8.5 and OpenMP 4.5.

\subsubsection{Dataset}
\label{sec:dataset}

The graphs employed in our experiments are detailed in Table \ref{tab:dataset}, obtained from the SuiteSparse Matrix Collection \cite{suite19}. These graphs encompass a range of $3.07$ to $214$ million vertices and $25.4$ million to $3.80$ billion edges. We make sure that the edges are both undirected and weighted, with a default weight set to $1$.

\input{src/tab-dataset}

\subsection{Results}

\ignore{It is observed that web graphs tend to have the highest inequality (Gini coefficient), while road networks tend to have the lowest.}As shown in Figure \ref{fig:im-all}, results indicate that the heuristics usually succeed in reducing inequality on graphs with high Gini coefficient (such as web graphs and social networks), but mostly fail on graphs with low Gini coefficient (such as road networks and collaboration networks). It is also observed that the rate of decrease in Gini coefficient decreases as more and more edges are added to graph. In general, we observe that the heuristics \textit{Cxrx}, \textit{CxSx}, and \textit{CxSr} perform the best, with \textit{CxSx}, and \textit{CxSr} performing almost identically. \textbf{Cxrx} and \textbf{CxSx} heuristics would therefore be the best choices, given that \textit{CxSr} requires a modified PageRank computation.

Based on these results, a suitable approach to minimizing inequality would be to apply both the \textit{Cxrx} and \textit{CxSx} heuristics and choose the the best among them for each edge addition. Future research work can include exploring randomized heuristics or looking for better deterministic heuristics.

\input{src/fig-im-all}

%% file: src/tab-dataset.tex
\begin{table}[!ht]
\centering
\caption{In our experiments, we use a list of 17 graphs. Each graph has its edges duplicated in the reverse direction to make them undirected, and a weight of 1 is assigned to each edge. The table lists the total number of vertices ($|V|$), total number of edges ($|E|$) after making the graph undirected, and the average degree of vertices ($D_{avg}$) for each graph. The number of vertices and edges are rounded to the nearest thousand or million, as appropriate.}
\label{tab:dataset}
\begin{tabular}{||c||c|c|c|c||}
  \toprule
  \textbf{Graph} &
  \textbf{$|V|$} &
  \textbf{$|E|$} &
  \textbf{$D_{avg}$} \\
  \midrule
  \multicolumn{4}{|c|}{\textbf{Web Graphs}} \\ \hline
    web-Stanford & 282K & 3.99M & 14.1 \\ \hline
    web-BerkStan & 685K & 13.3M & 19.4 \\ \hline
    web-Google & 916K & 8.64M & 9.43 \\ \hline
    web-NotreDame & 326K & 2.21M & 6.78 \\ \hline
    indochina-2004 & 7.41M & 304M & 41.0 \\ \hline
    \multicolumn{4}{|c|}{\textbf{Social Networks}} \\ \hline
    soc-Slashdot0811 & 77.4K & 1.02M & 13.2 \\ \hline
    soc-Slashdot0902 & 82.2K & 1.09M & 13.3 \\ \hline
    soc-Epinions1 & 75.9K & 811K & 10.7 \\ \hline
    soc-LiveJournal1 & 4.85M & 86.2M & 17.8 \\ \hline
    \multicolumn{4}{|c|}{\textbf{Collaboration Networks}} \\ \hline
    coAuthorsDBLP & 299K & 1.96M & 6.56 \\ \hline
    coAuthorsCiteseer & 227K & 1.63M & 7.18 \\ \hline
    coPapersCiteseer & 434K & 32.1M & 74.0 \\ \hline
    coPapersDBLP & 540K & 30.5M & 56.5 \\ \hline
    \multicolumn{4}{|c|}{\textbf{Road Networks}} \\ \hline
    italy\_osm & 6.69M & 14.0M & 2.09 \\ \hline
    great-britain\_osm & 7.73M & 16.3M & 2.11 \\ \hline
    germany\_osm & 11.5M & 24.7M & 2.15 \\ \hline
    asia\_osm & 12.0M & 25.4M & 2.12 \\ \hline
  \bottomrule
\end{tabular}
\end{table}

%% file: src/fig-im-all.tex
\begin{figure*}[hbtp]
  \centering
  \subfigure{
    \label{fig:im-key}
    \includegraphics[width=0.98\linewidth]{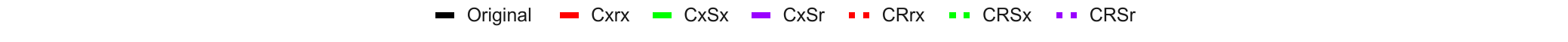}
  } \\[-2ex]
  \subfigure{
    \label{fig:im-web-Stanford}
    \includegraphics[width=0.23\linewidth]{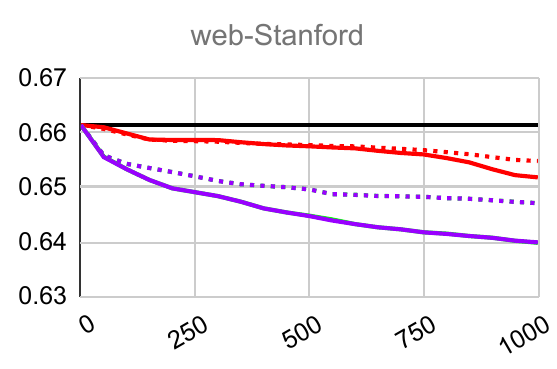}
  }
  \subfigure{
    \label{fig:im-web-BerkStan}
    \includegraphics[width=0.23\linewidth]{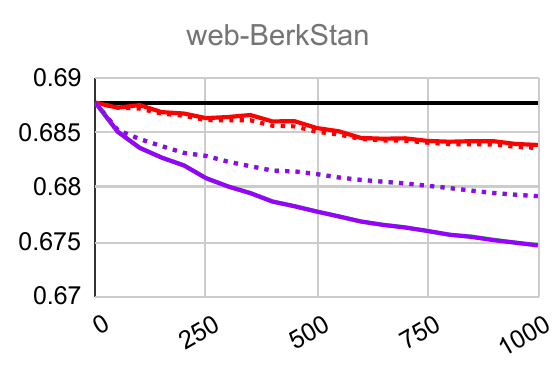}
  }
  \subfigure{
    \label{fig:im-web-Google}
    \includegraphics[width=0.23\linewidth]{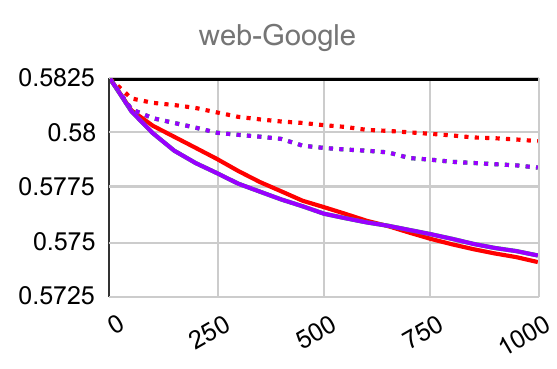}
  }
  \subfigure{
    \label{fig:im-web-NotreDame}
    \includegraphics[width=0.23\linewidth]{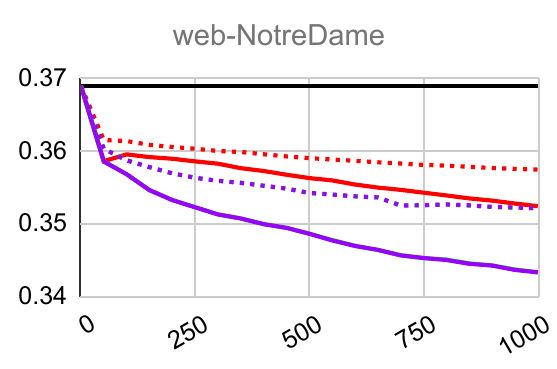}
  }
  \subfigure{
    \label{fig:im-soc-Slashdot0811}
    \includegraphics[width=0.23\linewidth]{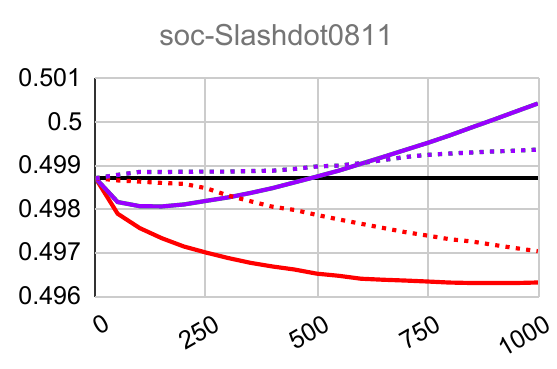}
  }
  \subfigure{
    \label{fig:im-soc-Slashdot0902}
    \includegraphics[width=0.23\linewidth]{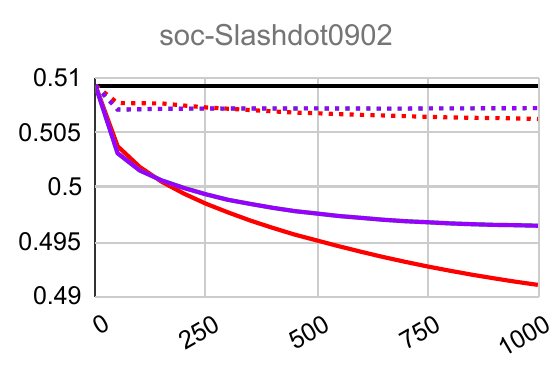}
  }
  \subfigure{
    \label{fig:im-soc-Epinions1}
    \includegraphics[width=0.23\linewidth]{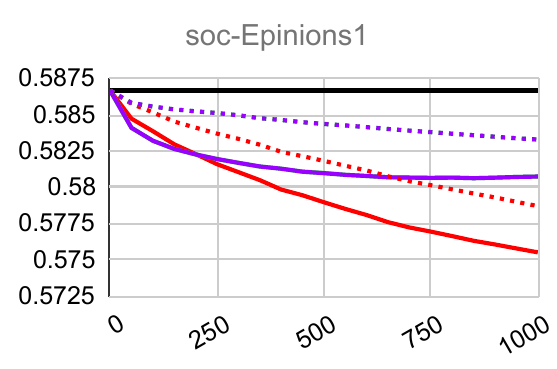}
  }
  \subfigure{
    \label{fig:im-soc-LiveJournal1}
    \includegraphics[width=0.23\linewidth]{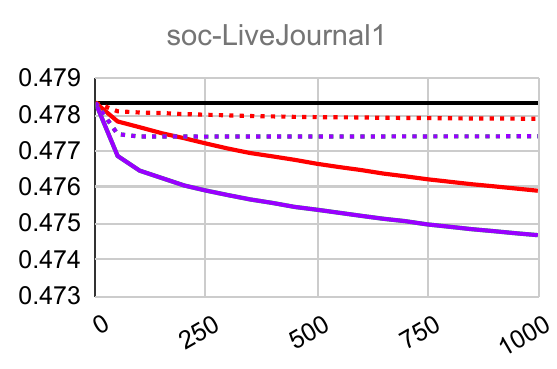}
  }
  \subfigure{
    \label{fig:im-coAuthorsDBLP}
    \includegraphics[width=0.23\linewidth]{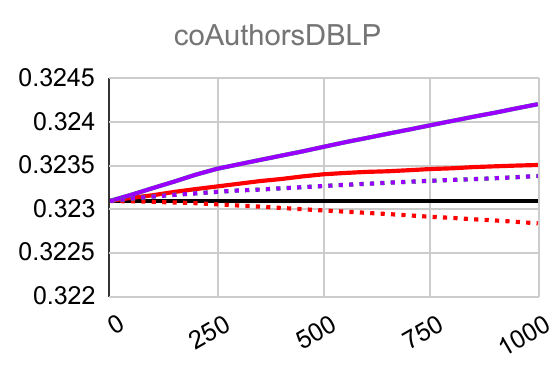}
  }
  \subfigure{
    \label{fig:im-coAuthorsCiteseer}
    \includegraphics[width=0.23\linewidth]{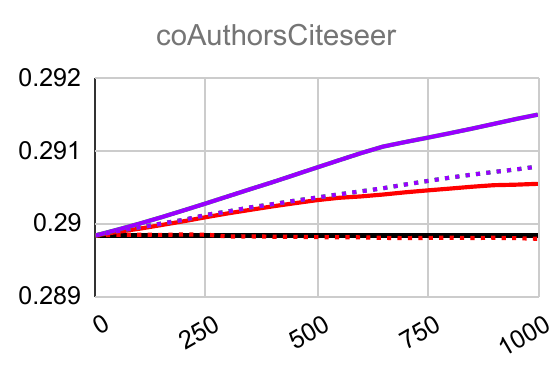}
  }
  \subfigure{
    \label{fig:im-coPapersCiteseer}
    \includegraphics[width=0.23\linewidth]{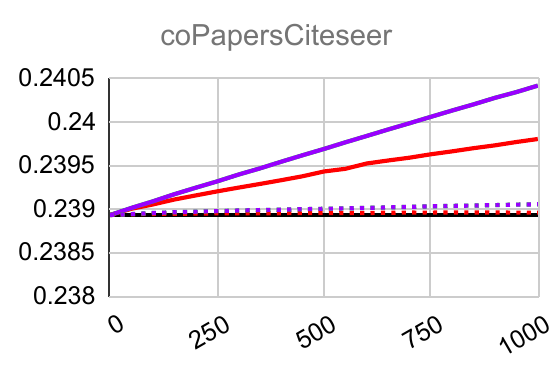}
  }
  \subfigure{
    \label{fig:im-coPapersDBLP}
    \includegraphics[width=0.23\linewidth]{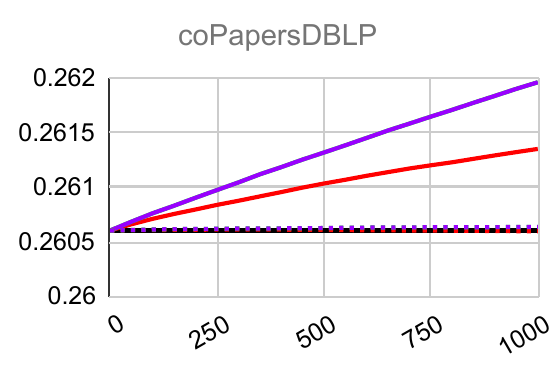}
  }
  \subfigure{
    \label{fig:im-italy_osm}
    \includegraphics[width=0.23\linewidth]{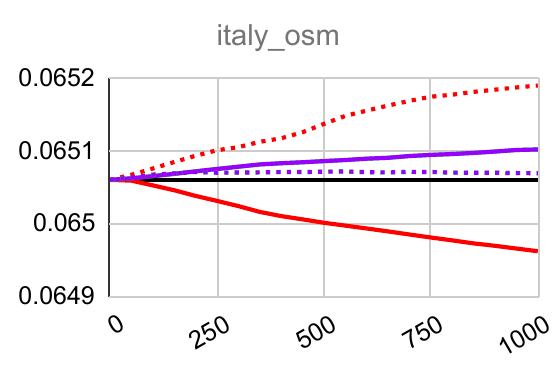}
  }
  \subfigure{
    \label{fig:im-great-britain_osm}
    \includegraphics[width=0.23\linewidth]{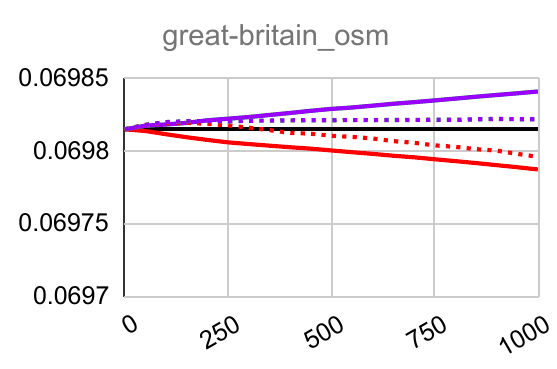}
  }
  \subfigure{
    \label{fig:im-germany_osm}
    \includegraphics[width=0.23\linewidth]{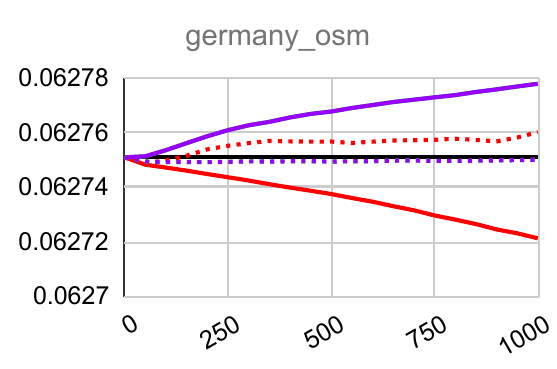}
  }
  \subfigure{
    \label{fig:im-asia_osm}
    \includegraphics[width=0.23\linewidth]{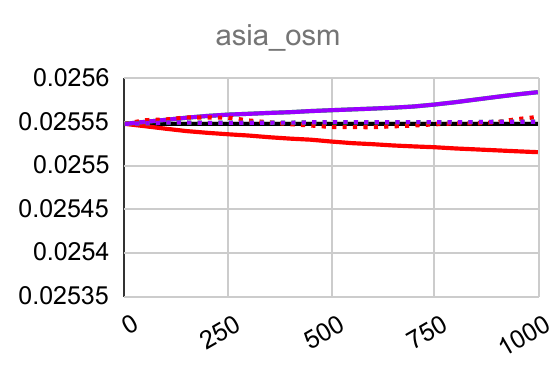}
  }
  \subfigure{
    \label{fig:im-indochina-2004}
    \includegraphics[width=0.23\linewidth]{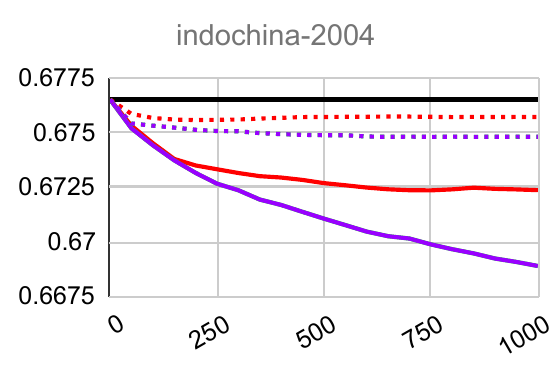}
  } \\[-2ex]
  \caption{Variation of Gini coefficient (Y-axis) with edges being added (X-axis) to the graphs incrementally with six different heuristics: \textit{Cxrx}, \textit{CxSx}, \textit{CxSr}, \textit{CRrx}, \textit{CRSx}, and \textit{CRSr} (see Algorithm \ref{alg:heuristic} for the psuedocode).}
  \label{fig:im-all}
\end{figure*}

%% file: 06-conclusion.tex
The study highlights that inequality is prevalent in web graphs, and demonstrates that efforts to minimize it are more effective in contexts with high Gini coefficients. The choice of heuristics plays a crucial role in reducing inequality. Our research suggests that a combination of \textbf{Cxrx} and \textbf{CxSx} heuristics may offer an effective approach to minimize inequality. Future research should continue to explore strategies to mitigate inequality in web ranking algorithms and promote a more equitable web environment.

%% file: main.bbl

\begin{thebibliography}{16}


\ifx \showCODEN    \undefined \def \showCODEN     #1{\unskip}     \fi
\ifx \showDOI      \undefined \def \showDOI       #1{#1}\fi
\ifx \showISBNx    \undefined \def \showISBNx     #1{\unskip}     \fi
\ifx \showISBNxiii \undefined \def \showISBNxiii  #1{\unskip}     \fi
\ifx \showISSN     \undefined \def \showISSN      #1{\unskip}     \fi
\ifx \showLCCN     \undefined \def \showLCCN      #1{\unskip}     \fi
\ifx \shownote     \undefined \def \shownote      #1{#1}          \fi
\ifx \showarticletitle \undefined \def \showarticletitle #1{#1}   \fi
\ifx \showURL      \undefined \def \showURL       {\relax}        \fi
\providecommand\bibfield[2]{#2}
\providecommand\bibinfo[2]{#2}
\providecommand\natexlab[1]{#1}
\providecommand\showeprint[2][]{arXiv:#2}

\bibitem[Dyer and Proll(1977)]%
        {dyer1977note}
\bibfield{author}{\bibinfo{person}{ME Dyer} {and} \bibinfo{person}{LG Proll}.} \bibinfo{year}{1977}\natexlab{}.
\newblock \showarticletitle{Note—on the validity of marginal analysis for allocating servers in m/m/c queues}.
\newblock \bibinfo{journal}{\emph{Management Science}} \bibinfo{volume}{23}, \bibinfo{number}{9} (\bibinfo{year}{1977}), \bibinfo{pages}{1019--1022}.
\newblock


\bibitem[Fox(1966)]%
        {fox1966discrete}
\bibfield{author}{\bibinfo{person}{Bennett Fox}.} \bibinfo{year}{1966}\natexlab{}.
\newblock \showarticletitle{Discrete optimization via marginal analysis}.
\newblock \bibinfo{journal}{\emph{Management science}} \bibinfo{volume}{13}, \bibinfo{number}{3} (\bibinfo{year}{1966}), \bibinfo{pages}{210--216}.
\newblock


\bibitem[Gini(1912)]%
        {gini1912italian}
\bibfield{author}{\bibinfo{person}{C Gini}.} \bibinfo{year}{1912}\natexlab{}.
\newblock \showarticletitle{Italian: Variabilit e mutabilit}.
\newblock \bibinfo{journal}{\emph{Variability and Mutability’, C. Cuppini, Bologna}} (\bibinfo{year}{1912}).
\newblock


\bibitem[Gutjahr and Fischer(2018)]%
        {gutjahr2018equity}
\bibfield{author}{\bibinfo{person}{Walter~J Gutjahr} {and} \bibinfo{person}{Sophie Fischer}.} \bibinfo{year}{2018}\natexlab{}.
\newblock \showarticletitle{Equity and deprivation costs in humanitarian logistics}.
\newblock \bibinfo{journal}{\emph{European Journal of Operational Research}} \bibinfo{volume}{270}, \bibinfo{number}{1} (\bibinfo{year}{2018}), \bibinfo{pages}{185--197}.
\newblock


\bibitem[Hung(2006)]%
        {hung2006allocation}
\bibfield{author}{\bibinfo{person}{Hui-Chih Hung}.} \bibinfo{year}{2006}\natexlab{}.
\newblock \emph{\bibinfo{title}{Allocation of jobs and resources to work centers}}.
\newblock \bibinfo{thesistype}{Ph.\,D. Dissertation}. \bibinfo{school}{The Ohio State University}.
\newblock


\bibitem[Karimi et~al\mbox{.}(2018)]%
        {karimi2018homophily}
\bibfield{author}{\bibinfo{person}{Fariba Karimi}, \bibinfo{person}{Mathieu G{\'e}nois}, \bibinfo{person}{Claudia Wagner}, \bibinfo{person}{Philipp Singer}, {and} \bibinfo{person}{Markus Strohmaier}.} \bibinfo{year}{2018}\natexlab{}.
\newblock \showarticletitle{Homophily influences ranking of minorities in social networks}.
\newblock \bibinfo{journal}{\emph{Scientific reports}} \bibinfo{volume}{8}, \bibinfo{number}{1} (\bibinfo{year}{2018}), \bibinfo{pages}{11077}.
\newblock


\bibitem[Kolodziej et~al\mbox{.}(2019)]%
        {suite19}
\bibfield{author}{\bibinfo{person}{S. Kolodziej}, \bibinfo{person}{M. Aznaveh}, \bibinfo{person}{M. Bullock}, \bibinfo{person}{J. David}, \bibinfo{person}{T. Davis}, \bibinfo{person}{M. Henderson}, \bibinfo{person}{Y. Hu}, {and} \bibinfo{person}{R. Sandstrom}.} \bibinfo{year}{2019}\natexlab{}.
\newblock \showarticletitle{{The SuiteSparse matrix collection website interface}}.
\newblock \bibinfo{journal}{\emph{The Journal of Open Source Software}} \bibinfo{volume}{4}, \bibinfo{number}{35} (\bibinfo{date}{Mar} \bibinfo{year}{2019}), \bibinfo{pages}{1244}.
\newblock


\bibitem[Krasanakis et~al\mbox{.}(2021)]%
        {krasanakis2021applying}
\bibfield{author}{\bibinfo{person}{E. Krasanakis}, \bibinfo{person}{S. Papadopoulos}, {and} \bibinfo{person}{I. Kompatsiaris}.} \bibinfo{year}{2021}\natexlab{}.
\newblock \showarticletitle{Applying fairness constraints on graph node ranks under personalization bias}. In \bibinfo{booktitle}{\emph{Complex Networks \& Their Applications IX: Volume 2, Proceedings of the Ninth International Conference on Complex Networks and Their Applications COMPLEX NETWORKS 2020}}. Springer, \bibinfo{pages}{610--622}.
\newblock


\bibitem[Page et~al\mbox{.}(1999)]%
        {rank-page99}
\bibfield{author}{\bibinfo{person}{L. Page}, \bibinfo{person}{S. Brin}, \bibinfo{person}{R. Motwani}, {and} \bibinfo{person}{T. Winograd}.} \bibinfo{year}{1999}\natexlab{}.
\newblock \bibinfo{booktitle}{\emph{{The PageRank citation ranking: Bringing order to the web.}}}
\newblock \bibinfo{type}{{T}echnical {R}eport}. \bibinfo{institution}{Stanford InfoLab}.
\newblock


\bibitem[Pitoura(2023)]%
        {pitoura2023pagerank}
\bibfield{author}{\bibinfo{person}{E. Pitoura}.} \bibinfo{year}{2023}\natexlab{}.
\newblock \showarticletitle{Pagerank Fairness in Networks}.
\newblock  (\bibinfo{year}{2023}).
\newblock


\bibitem[Saxena et~al\mbox{.}(2022)]%
        {saxena2022fairsna}
\bibfield{author}{\bibinfo{person}{A. Saxena}, \bibinfo{person}{G. Fletcher}, {and} \bibinfo{person}{M. Pechenizkiy}.} \bibinfo{year}{2022}\natexlab{}.
\newblock \showarticletitle{Fairsna: Algorithmic fairness in social network analysis}.
\newblock \bibinfo{journal}{\emph{arXiv preprint arXiv:2209.01678}} (\bibinfo{year}{2022}).
\newblock


\bibitem[Shehadeh and Snyder(2023)]%
        {shehadeh2023equity}
\bibfield{author}{\bibinfo{person}{Karmel~S Shehadeh} {and} \bibinfo{person}{Lawrence~V Snyder}.} \bibinfo{year}{2023}\natexlab{}.
\newblock \showarticletitle{Equity in stochastic healthcare facility location}.
\newblock In \bibinfo{booktitle}{\emph{Uncertainty in Facility Location Problems}}. \bibinfo{publisher}{Springer}, \bibinfo{pages}{303--334}.
\newblock


\bibitem[Tsioutsiouliklis et~al\mbox{.}(2022)]%
        {tsioutsiouliklis2022link}
\bibfield{author}{\bibinfo{person}{S. Tsioutsiouliklis}, \bibinfo{person}{E. Pitoura}, \bibinfo{person}{K. Semertzidis}, {and} \bibinfo{person}{P. Tsaparas}.} \bibinfo{year}{2022}\natexlab{}.
\newblock \showarticletitle{Link Recommendations for PageRank Fairness}. In \bibinfo{booktitle}{\emph{Proceedings of the ACM Web Conference 2022}}. \bibinfo{pages}{3541--3551}.
\newblock


\bibitem[Tsioutsiouliklis et~al\mbox{.}(2021)]%
        {tsioutsiouliklis2021fairness}
\bibfield{author}{\bibinfo{person}{S. Tsioutsiouliklis}, \bibinfo{person}{E. Pitoura}, \bibinfo{person}{P. Tsaparas}, \bibinfo{person}{I. Kleftakis}, {and} \bibinfo{person}{N. Mamoulis}.} \bibinfo{year}{2021}\natexlab{}.
\newblock \showarticletitle{Fairness-aware pagerank}. In \bibinfo{booktitle}{\emph{Proceedings of the Web Conference 2021}}. \bibinfo{pages}{3815--3826}.
\newblock


\bibitem[Xie et~al\mbox{.}(2021)]%
        {xie2021fairrankvis}
\bibfield{author}{\bibinfo{person}{T. Xie}, \bibinfo{person}{Y. Ma}, \bibinfo{person}{J. Kang}, \bibinfo{person}{Ha. Tong}, {and} \bibinfo{person}{R. Maciejewski}.} \bibinfo{year}{2021}\natexlab{}.
\newblock \showarticletitle{Fairrankvis: A visual analytics framework for exploring algorithmic fairness in graph mining models}.
\newblock \bibinfo{journal}{\emph{IEEE Transactions on Visualization and Computer Graphics}} \bibinfo{volume}{28}, \bibinfo{number}{1} (\bibinfo{year}{2021}), \bibinfo{pages}{368--377}.
\newblock


\bibitem[Yang et~al\mbox{.}(2013)]%
        {yang2013call}
\bibfield{author}{\bibinfo{person}{Muer Yang}, \bibinfo{person}{Theodore~T Allen}, \bibinfo{person}{Michael~J Fry}, {and} \bibinfo{person}{W~David Kelton}.} \bibinfo{year}{2013}\natexlab{}.
\newblock \showarticletitle{The call for equity: simulation optimization models to minimize the range of waiting times}.
\newblock \bibinfo{journal}{\emph{IIE Transactions}} \bibinfo{volume}{45}, \bibinfo{number}{7} (\bibinfo{year}{2013}), \bibinfo{pages}{781--795}.
\newblock


\end{thebibliography}
